\documentclass[twocolumn,aps,prb,showpacs,groupedaddress,amsfonts,amssymb,amsmath]{revtex4-2}
\usepackage{lipsum}
\usepackage{graphicx}
\usepackage{subfigure}
\usepackage{epstopdf}
\usepackage{makecell}
\usepackage{booktabs}
\usepackage{siunitx}
\usepackage{microtype}
\usepackage{url}
\usepackage[colorlinks=true, letterpaper=true, pdfstartview=FitV, linkcolor=blue, citecolor=blue, urlcolor=blue]{hyperref}

\begin{document}
\draft

\title{The fate of reentrant localization phenomenon in the one-dimensional dimerized quasiperiodic chain with long-range hopping}

\author{Haoyu Wang,$^{1,4}$ Xiaohong Zheng,$^2$ Jun Chen,$^{3,4,*}$ Liantuan Xiao,$^{1,4}$ Suotang Jia,$^{1,4}$ Lei Zhang,$^{1,4,\dagger}$}

\address{$^1$State Key Laboratory of Quantum Optics and Quantum Optics Devices, Institute of Laser Spectroscopy, Shanxi University, Taiyuan 030006, China\\
$^2$College of Information Science and Technology, Nanjing Forestry University, Nanjing 210037, China\\
$^3$State Key Laboratory of Quantum Optics and Quantum Optics Devices, Institute of Theoretical Physics, Shanxi University, Taiyuan 030006, China\\
$^4$Collaborative Innovation Center of Extreme Optics, Shanxi University, Taiyuan 030006, China}

\begin{abstract}
Recently, the exciting reentrant localization transition phenomenon was found in a one-dimensional dimerized lattice with staggered quasiperiodic potentials. Usually, long-range hopping is typically important in actual physical systems. In this work, we study the effect of next-nearest neighbor hopping (NNNH) on the reentrant localization phenomenon. Due to the presence of NNNH, the broken chiral symmetry is further enhanced and the localization properties of electron states in the upper and lower bands become quite different. It is found that the reentrant localization can still persist within a range of NNNH both in Hermitian and non-Hermitian cases. Eventually, the reentrant localization disappears as the strength of NNNH increases to some extent, since the increasing NNNH weakens the dimerization of the system and destroys its competition with the quasiperiodic disorder. Our work thus reveals the effect of long-range hopping in the reentrant localization phenomenon and deepens its physical understanding.
\end{abstract}

\maketitle

\section{Introduction} \label{Sec:I}
As an important research topic in condensed matter physics, Anderson localization has been extensively studied since the pioneering work of Anderson \cite{lee1985disordered,evers2008anderson}.
In recent years, the phenomenon of quantum particle localization which is directly related to their transport properties has attracted extensive attention.
Anderson localization describes the absence of electron diffusion of electronic waves aroused by disorder and predicts the metal-insulator transition due to quantum interference of scattered electron wave functions\cite{anderson1958absence,roati2008anderson,lahini2009observation}, which sparked widespread research interest in many systems, such as cold atomic gases \cite{billy2008direct, roati2008anderson, kondov2011three, jendrzejewski2012three, semeghini2015measurement, pasek2017anderson, hainaut2019experimental, richard2019elastic}, quantum optics \cite{sperling2013direct, wiersma1997localization, storzer2006observation, schwartz2007transport, lahini2008anderson}, acoustic waves systems \cite{hu2008localization}, etc. According to the scaling theory \cite{abrahams1979scaling} of Anderson localization, all the single-particle states will be spatially exponentially localized by arbitrarily small random (uncorrelated) disorders in one- and two-dimensional systems. While in the three-dimension, both extended and localized states can coexist in the system with disorders and an energy-dependent mobility edge distinguishing the localized region from the delocalized region that appears at the phase boundary. Compared to randomly disordered lattices, quasiperiodic systems are located at the interface between long-range ordered and disordered systems, which provides a unique opportunity to explore localization transitions. The most typical example is the Aubry-Andr\'e –Harper (AAH) model \cite{aubry1980analyticity,harper1955single}, which hosts an energy-independent localization transition and has been widely investigated in optical and atomic systems \cite{lahini2009observation, kraus2012topological, roati2008anderson, xue2014observation}. In this model, all eigenstates change from extended to localized over a critical quasiperiodic amplitude because of its self-dual symmetry \cite{aubry1980analyticity, suslov1982localization, wilkinson1984critical}.
However, many generalized AAH models \cite{sarma1986proposed, biddle2010predicted, ganeshan2015, sun2015localization, gopalakrishnan2017self, purkayastha2017nonequilibrium} exhibit accurate energy-dependent mobility edges \cite{biddle2010predicted, luschen2018single, wang2020one, ganeshan2015, PhysRevB.101.014205, PhysRevLett.125.200604, PhysRevLett.123.070405, PhysRevLett.125.060401, PhysRevB.96.085119} when long-range hopping terms \cite{deng2019one, biddle2011localization, gopalakrishnan2017self, liu2018mobility, deng2019one, saha2019anomalous, PhysRevB.96.085119, boers2007mobility, diener2001transition} or modified quasiperiodic potentials are introduced. These accurate mobility edges are helpful for a better understanding of Anderson localization in one-dimensional quasicrystals.

Lately, the interplay of non-Hermiticity and disorder has received a lot of research attention. For a non-Hermitian system, the non-Hermiticity can be generally obtained by introducing nonreciprocal hopping terms or gain and loss potentials, which are found in open systems exchanging energy or particles with the environment. Consequently, lots of exotic phenomena have been found that not exist in the traditional Hermitian systems, such as non-Hermitian skin effect \cite{yao2018edge, kunst2018biorthogonal, alvarez2018non, lee2019anatomy, yokomizo2019non, zhang2020correspondence, okuma2020topological, li2020critical, xiao2020non}, exceptional points \cite{berry2004physics, heiss2012physics, miri2019exceptional, regensburger2012parity, miri2019exceptional}, and exotic transport features \cite{longhi2020non, yi2020non, liu2020helical, lee2020unraveling, longhi2015non, longhi2015robust, jin2017one}. On the spectral side, if the system is in the $\mathcal{PT}$ symmetry phase \cite{longhi2019topological, longhi2019metal, jiang2019interplay}, the energy spectrum can still be purely real. However, when the non-Hermitian parameter exceeds the exceptional points, the $\mathcal{PT}$ symmetry is broken, resulting a real-to-complex transition in the energy spectrum. Which have attracted widespread research attention in the fields of topolectrical (TE) circuits \cite{PhysRevB.101.020201, PhysRevResearch.2.033052, Ashida_2020, PhysRevA.84.040101, PhysRevLett.122.247702, Rafi-Ul-Islam_2021, helbig2020generalized}, acoustics \cite{ma2016acoustic, cummer2016controlling, zangeneh2019active}, ultracold atoms \cite{daley2014quantum, kuhr2016quantum, PhysRevB.101.014205, PhysRevLett.125.200604, PhysRevLett.123.070405, PhysRevLett.125.060401, PhysRevB.96.085119}, disordered systems \cite{beenakker1997random, fyodorov2003random, mitchell2010random}, etc.

More recently, a interesting reentrant localization phenomenon in one-dimensional quasiperiodic disordered systems has been found and studied in both Hermitian and non-Hermitian cases \cite{roy2021reentrant, wu2021non, jiang2021mobility}. This non-trivial reentrant feature and corresponding single-particle mobility edge (SPME) can be attributed to the competition between the hopping dimerization and the staggered disorder. During this process, some already localized states become extended again as the strength of staggered quasi-periodic potential increases. This result in two localization transitions until all the states are finally completely localized. Usually, long-range hopping plays a significant role in the actual physical systems\cite{di2014ultracold, viyuela2018chiral, degottardi2013majorana}. For example, long-range hopping is important to generate the mobility edge and Anderson localization\cite{deng2019one,roy2021fraction}. Then the natural question arises: what is the fate of the reentrant localization transition in a 1D quasiperiodic system with long-range hopping? To simplify the model and facilitate research, the next-nearest neighbor hopping(NNNH) is introduced in three kinds of generalized AAH chains with staggered quasiperiodic potential. In this work, we showed that increasing NNNH impairs the dimerization of the SSH chain and destroys its competition with the quasi-periodic disorder. When the strength of NNNH reaches some value, the reentrant phenomenon disappears. To confirm our findings, we investigated the eigenenergy spectrum, and participation ratios in this work, and showed that the introduction of long-range interactions has a great impact on the reentrant localization transitions.

This paper is organized as follows. In Sec. \ref{Sec:II}, we introduced the model based on a generalized AAH quasiperiodic chain and calculated the corresponding significant physical quantities that can be used to unveil the spectral, localization, topological transitions, and mobility edges in the quasiperiodic systems. The numerical results and discussions are presented in Sec. \ref{Sec:III}, in which we calculate the eigenenergy spectrum, participation ratios, for three different cases to confirm our findings. Finally, a brief summary is given in Sec. \ref{Sec:IV}.
\section{model and approach} \label{Sec:II}
\begin{figure}[]
\includegraphics[trim=0mm 0mm 0mm 0mm,scale=0.375]{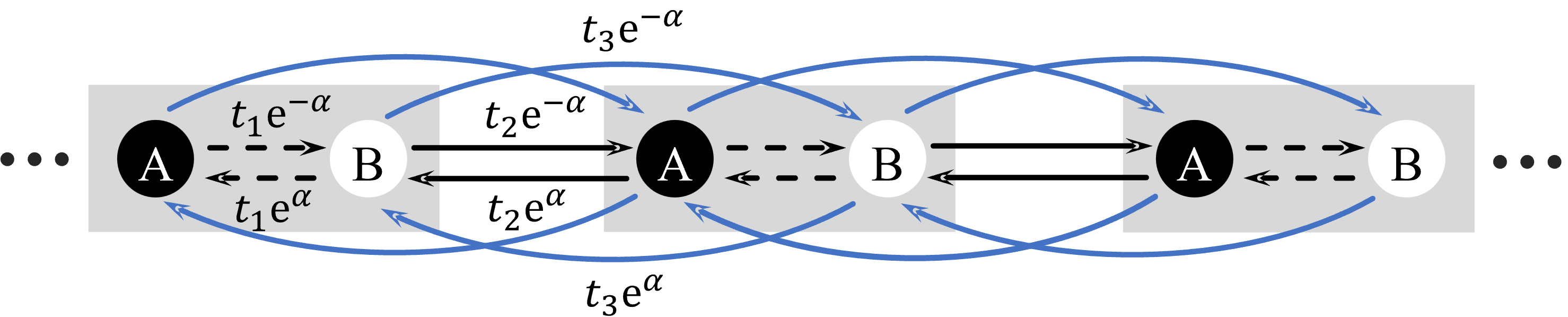}
\caption{Schematic plot of the modulated AAH model with next-nearest neighbor hopping terms, each unit cell consists of two sublattices \emph{A} and \emph{B}. $t_1$, $t_{2}$ and $t_3$ represent intra-cell, inter-cell and next-nearest neighbor hopping strengths, respectively. Parameter $\alpha$ denotes the asymmetric hopping strength.}\label{Fig1}
\end{figure}
We study the following one-dimensional tight-binding model with on-site quasiperiodic potential as shown in Fig. \ref{Fig1},
\begin{widetext}
\begin{equation}	\label{eqI}
\begin{gathered}
	\hat{H}=t_{1}\sum_{n=1}^{N}\left ( e^{\alpha}\hat{c}_{n,B}^{\dagger} \hat{c}_{n,A} +e^{-\alpha }\hat{c}_{n,A}^{\dagger} \hat{c}_{n,B}\right )+t_{2}\sum_{n=1}^{N-1}\left (e^{\alpha}\hat{c}_{n+1,A}^{\dagger}\hat{c}_{n,B}+e^{-\alpha}\hat{c}_{n,B}^{\dagger} \hat{c}_{n+1,A}\right )  \\
	+t_{3} \sum_{n=1}^{N-2}\left(e^{\alpha} \hat{c}_{n+1, A}^{\dagger} \hat{c}_{n, A}+e^{-\alpha} \hat{c}_{n, A}^{\dagger} \hat{c}_{n+1, A}+e^{\alpha} \hat{c}_{n+1, B}^{\dagger} \hat{c}_{n, B}+e^{-\alpha} \hat{c}_{n, B}^{\dagger} \hat{c}_{n+1, B}\right) \\
	+\sum_{n} V_{A}\hat{n}_{n,A}cos \left [ 2\pi \beta (2n-1) +i\Delta \right ] +
	\sum_{n} V_{B}\hat{n}_{n,B}cos \left [ 2\pi \beta (2n) +i\Delta \right ]
\end{gathered}
\end{equation}
\end{widetext}
where $n$ represents the unit cell index and the length of chain is $L=2N$, $ \hat{c}^{\dagger}_{n,A/B}$ ($ \hat{c}_{n,A/B}$) are creation (annihilation) operators corresponding to \emph{A} or \emph{B} sublattice denoted by $(n,A)$ and $(n,B)$. The particle number operators on related sites are $\hat{n}_{n,A}$ and $\hat{n}_{n,B}$. Here, intra- and intercell hopping strengths are represented by $t_1$, $t_2$, and $t_3$ represents the next-nearest neighbor hopping strength. Parameters $V_{A}$ and $V_{B}$ are the strength of the on-site quasiperiodic potentials at sublattices $A$ and $B$. $\beta$ determines the period of quasiperiodic potential. Here we introduce the staggered potential in the sublattice by assuming $V_{A}=-V_{B}=V$. It should be noted that the nonreciprocal strength $\alpha$ and the complex phase factor $i\Delta$ contribute the non-Hermiticity in this model. In the quasiperiodic potential term, we take the irrational number defined as $\beta =\lim_{m \to \infty} (\frac{F_{m-1}}{F_m})$ , where $ \left \{ F_{m} \right \} $ are the Fibonacci numbers \cite{kohmoto1983metal, wang2016spectral}. In our work, we chose $\beta=(\sqrt{5}-1)/2$ as a Diophantine number \cite{jitomirskaya1999metal}. Specifically, for most case we set $L=610, t_{1}=1$ and the periodic boundary condition (PBC) unless otherwise mentioned. In the following, the reentrant localization phenomenon in three different cases are studied, which are listed in Table \ref{I}. In subsection \ref{Sec:III A}, a Hermitian system with $\alpha=0, \Delta=0$ is studied. In subsections \ref{Sec:III B} and \ref{Sec:III C}, the numerical results and discussion of non-Hermitian systems with nonzero asymmetric hopping strength $\alpha$ or complex on-site potential $i\Delta$ are presented, respectively.

\begin{table}[]
\caption{\label{tab:fonts} The different nonreciprocal and complex phase factors on reentrant localization transitions is discussed for three Hamiltonian cases, i.e., M1, M2, and M3. $\alpha$ is the nonreciprocal strength for asymmetric hopping terms. $i\Delta$ donates the complex phase factor in the quasiperiodic potential.}
	\begin{tabular*}{8.6cm}{c@{\extracolsep{\fill}}ccr}
		\toprule
		\toprule
		Model & nonreciprocal strength & complex phase factor &\\
		\hline
		M1    &        $\alpha=0$         &   $\Delta=0$    & \\
		M2    &        $\alpha\neq0$        &   $\Delta=0$    & \\
		M3    &        $\alpha=0$         &   $\Delta\neq0$    & \\
		\bottomrule
		\bottomrule
		\label{I}
	\end{tabular*}
\end{table}

In most cases, the inverse participation ratio (IPR) and normalized participation ratio (NPR) \cite{PhysRevB.96.085119, li2020mobility} are used to identify localized and extended states in the system, which are defined as
\begin{align}
	\mathrm{IPR} ^{i} & =\frac{\sum_{n=1}^{L} \left | \psi_{n}^{i} \right | ^4}{\left[\sum_{n=1}^{L}\left|\psi_{n}^{i} \right| ^{2} \right]^2}, \\
	\mathrm{NPR} ^{i} & =\left[L\sum_{n=1}^{L}\left |\psi_{n}^{i} \right |^{4} \right]^{-1},
\end{align}
where $\psi_{n}^{i}$ is the eigenstate with the superscript $i$ denoting $i$th eigenstate, and $n$ denotes the lattice  	 .

Furthermore, by taking summation over all eigenstates, we can obtain the IPR and NPR of the system
\begin{align}
	\left \langle \mathrm{IPR} \right \rangle & =\frac{1}{L}\sum_{i=1}^{L}\mathrm{IPR} ^{i}, \\
	\left \langle \mathrm{NPR} \right \rangle & =\frac{1}{L}\sum_{i=1}^{L}\mathrm{NPR} ^{i},
\end{align}
and $\eta$\cite{li2020mobility}
\begin{equation}
	\eta = \log_{10}{[\left \langle \mathrm{IPR} \right \rangle \left \langle \mathrm{NPR} \right \rangle]},
\end{equation}
to figure out whether the system is in the intermediate, fully extended or localized phases. For extended states, the $\left\langle\mathrm{IPR}\right\rangle$ tends to be zero (finite) and the $\left\langle\mathrm{NPR}\right\rangle$ tends to be finite (zero) in the large $L$ limit (localized states). Based on these, the localization transitions can be identified. For convenience, the typical orders of $\left\langle\mathrm{IPR}\right\rangle$ and $\left\langle\mathrm{NPR}\right\rangle$ in different regions have been listed in Table \ref{II}.

In order to more clearly exhibit the SPME, the fractal dimension $\Gamma^i$ of the wave function is introduced as the probe of the $i$th wavefunction's localization character\cite{xia2021new, evers2008anderson, PhysRevB.96.085119},
\begin{equation}
	\Gamma^i=-\frac{\ln \left ( \sum_{n=1}^{L}\left|\psi_{n}^{i}\right|^{4} \right ) }{\ln L}.
\end{equation}
Contrary to the $\mathrm{IPR}^{i}$ mentioned above, the $\Gamma^i$ tends to be zero for localized states and finite for extended states.

\begin{table}
	\caption{\label{tab:fonts} A convenient operational definition to distinguish three different localization phases in a 1D single-particle Hamiltonian. Here $L$ is the size of the system.}
	\begin{tabular*}{8.6cm}{l@{\extracolsep{\fill}}c}
		\toprule
		\toprule
		Localized phase    & $\langle\mathrm{IPR}\rangle \sim \mathcal{O}(1) \text { and }\langle\mathrm{NPR}\rangle \sim L^{-1}$ \\
		Extended phase     & $\langle\mathrm{IPR}\rangle \sim L^{-1} \text { and }\langle\mathrm{NPR}\rangle \sim \mathcal{O}(1)$ \\
		Intermediate phase & $\langle\mathrm{IPR}\rangle \sim \mathcal{O}(1) \text { and }\langle\mathrm{NPR}\rangle \sim \mathcal{O}(1) $ \\
		\bottomrule
		\bottomrule
		\label{II}
	\end{tabular*}
\end{table}

\section{numerical results and discussions} \label{Sec:III}

\subsection{M1: Hermitian Hamiltonian} \label{Sec:III A}
First of all, we study the system described by hermitian Hamiltonian M1 without considering the effect of nonreciprocal strength and complex phase factor as shown in Table \ref{I}.
\begin{figure}[]
\centering
\includegraphics[trim=5mm 0mm 0mm 0mm,scale=0.41]{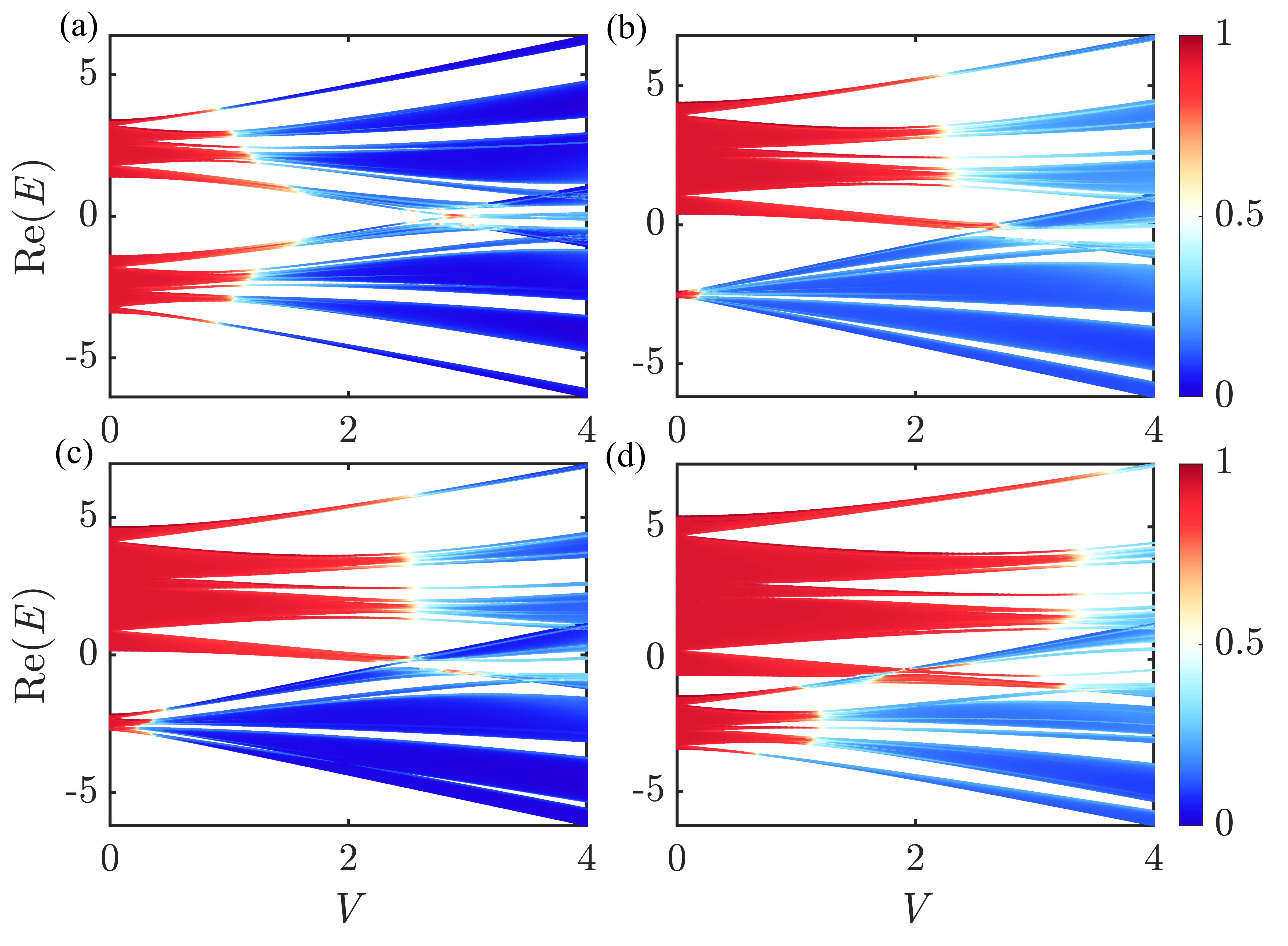}
\caption{The energy spectrum versus quasiperiodic modulation amplitude $V$ for the system with $t_1=1,t_2=2.4$. (a) $t_ 3=0$, (b) $t_ 3=0.5$, (c) $t_ 3=0.62$, (d) $t_3=1$	. The colormap shows $\Gamma^i$ associated with $i$th eigenstate.}
\label{Fig2}
\end{figure}
After diagonalization, the energy spectrums with corresponding $\Gamma^i$ are calculated and presented in Fig. \ref{Fig2}. When there is no next-nearest neighbor hopping (NNNH), the energy spectrum given in Fig. \ref{Fig2}(a). Since $\Gamma^i$ changes from $1$ to 0 as the quasiperiodic potential $V$ increases from 0 around 2.5, we can know that all extended states are gradually localized. There exists a critically intermediate region with the coexistence of both extended and localized states separated by the SPME. As $V$ further continues to increase, the upper bands and lower bands touch and some localized states become extended again with $V\simeq 2.5$. Thus another critical region with SPME emerges, which is dubbed as the reentrant phenomenon discovered recently\cite{roy2021reentrant}. However, by introducing the next-nearest neighbor hopping shown in Fig. \ref{Fig1}, the broken chiral symmetry is further enhanced, making the eigenenergy spectrum asymmetric up and down. From Fig. \ref{Fig2}(b), we can know that the extended states of lower bands tend to be easier localized than that of the upper bands as $V$ increases for $t_3=0.5$. More importantly, the states of the lowest branch in upper bands persist to be extended until $V$ reaches the second critical region as the strength of NNNH increases. Thus, there is only one critical region with SPME left, which implies the disappearance of the reentrant phenomenon with the introduction of long-range hopping. Interestingly, it is found that the reentrant phenomenon reappears when $V\simeq 2.8$ and $t_3$ increases up to $0.62$ in Fig.\ref{Fig2} (c). Clearly, as $t_3$ further increases up to $1$, there is no signature of the reentrant phenomenon in Fig.\ref{Fig2} (d). From the above, we know that the reentrant phenomena can be eliminated by introducing next-nearest hopping. However, this is not always the case, and further investigations are needed.

\begin{figure}[]
\centering 	
\includegraphics[trim=5mm 0mm 0mm 0mm,scale=0.36]{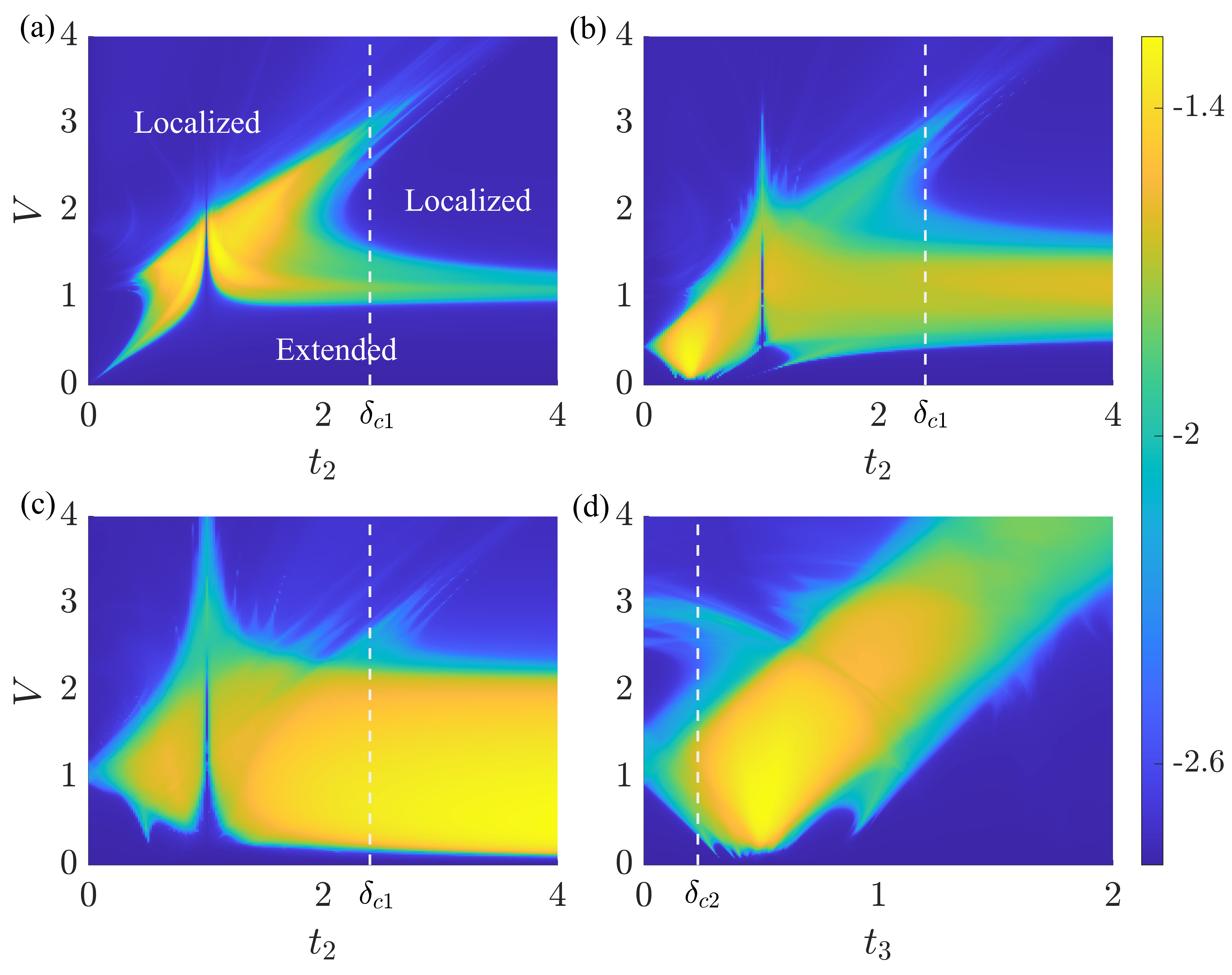}
\caption{Phase diagrams of the Hermitian system in $V$ and $t_2$ plane for (a) $t_3=0$, (b) $t_3=0.2$, (c) $t_3=0.5$. Phase diagrams of the Hermitian system for in $V$ and $t_3$ plane for (d) $t_2=2.4$. Colorbar represents the value of $\eta$. Note that the extended and localized phases are represented by dark blue regions, and the brilliant yellow region shows the intermediate phase with SPME.}\label{Fig3}
\end{figure}
The quantity $\eta$ is usually used to distinguish the critically intermediated region from fully extended and localized regions. From it, one can easily know how many times the system enters the intermediated regions to confirm the reentrant phenomena that are found with energy spectrum calculations. Therefore, we present the $\eta$ in form of a phase diagram as shown in Fig. \ref{Fig3}. First of all, we can find that all eigenstates are extended when the quasiperiodic potential is small. As the staggered potential $V$ increases, some originally extended states are localized, forming a critical region where extended and local states coexist in the phase diagram in $V$ and $t_2$ plane without considering NNNH in Fig. \ref{Fig3}(a). One can encounter the critical regime twice for a range of $t_2=[2.2, 3]$, which is indeed a reentrant phenomenon. From Fig. \ref{Fig3}(b,c), we can know that the range of $t_2$ presenting reentrant phenomenon shrinks as the increase of NNNH. In particular, the reentrant phenomenon with $t_2=2.4$ disappear when $t_3 \ge 0.2$, which is consistent with energy spectrum analysis. To clarify the effect of NNNH, we plot the phase diagram in the $V$-$t_3$ plane in Fig. \ref{Fig3}(d). As the NNNH increases, it can be seen that two critical regions merge beyond a critical point $\delta_{c2} \approx 0.23$. This confirm what we find in the energy spectrum calculations. In the meanwhile, the additional lobe of the intermediate phase appears again and soon fades away in some narrow parameter intervals for $t_3\in (0.56,0.66)$, which is also shown in the energy spectrum calculations. As can be seen, the reentrant feature vanished eventually in the one-dimensional staggered quasiperiodic generalized AAH lattice with the introduction of long-range hopping.

\begin{figure}[]
	\centering 	
	\includegraphics[trim=5mm 0mm 0mm 0mm,scale=0.42]{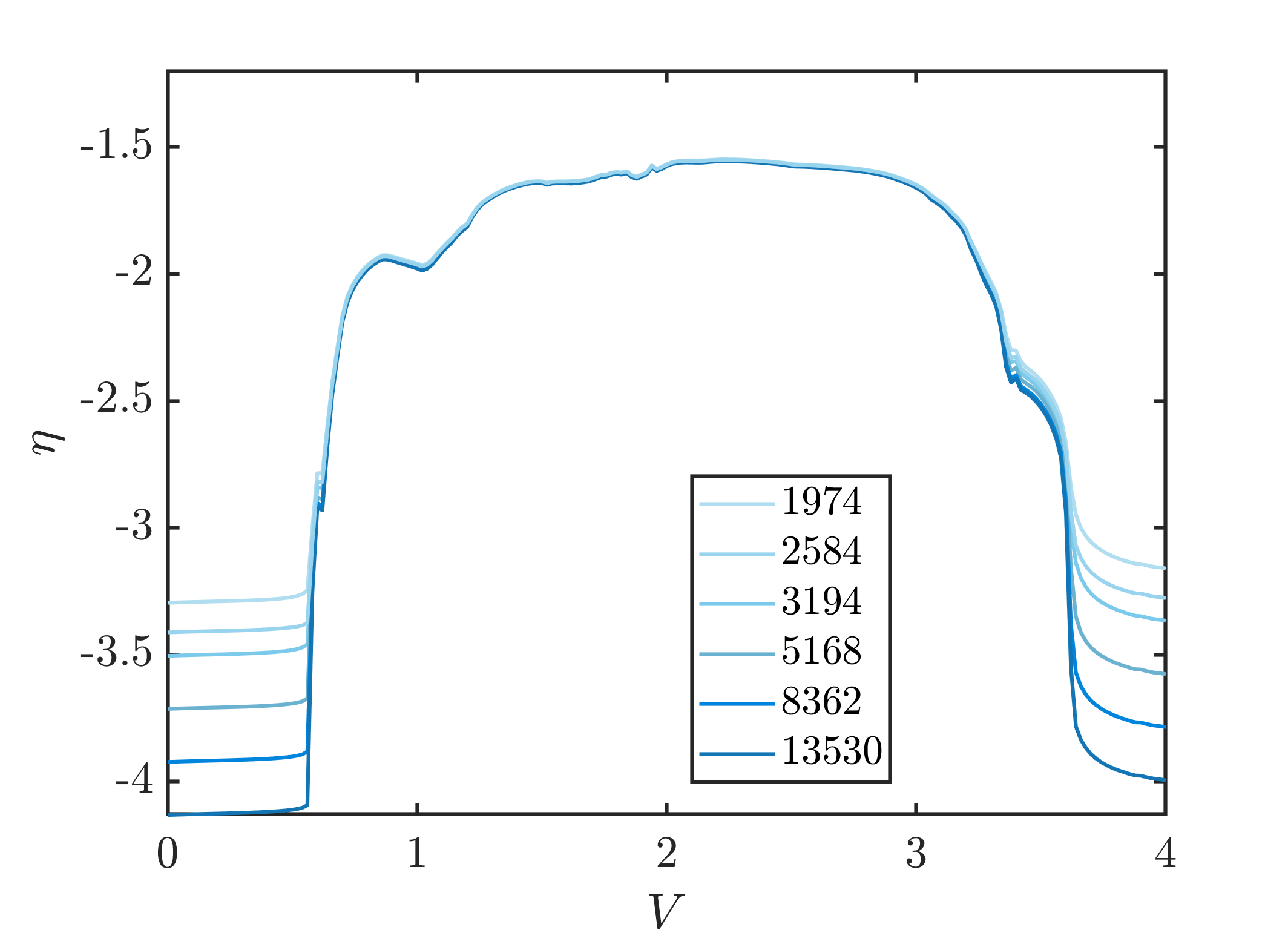}
	\caption{$\eta$ versus $V$ for different system sizes $L= 1974,2584,3194,5158,8362,13530$ with $t_1=1, t_2=2.4, t_3=1$.}\label{Fig4}
\end{figure}
To further verify the effect of NNNH on the disappearance of reentrant localization, we perform finite size calculations for different system sizes when NNNH is fixed. It is found that $\eta$ decreases as the system size increases in both extend and localized regions in Fig. \ref{Fig4}. Instead, the only one intermediate region is more notable. This clearly indicates that the stability of NNNH effect on reentrant phenomenon.

\subsection{M2: non-Hermitian Hamiltonian with asymmetric couplings} \label{Sec:III B}
\begin{figure}[]
	\centering
	\includegraphics[trim=5mm 0mm 0mm 0mm,scale=0.44]{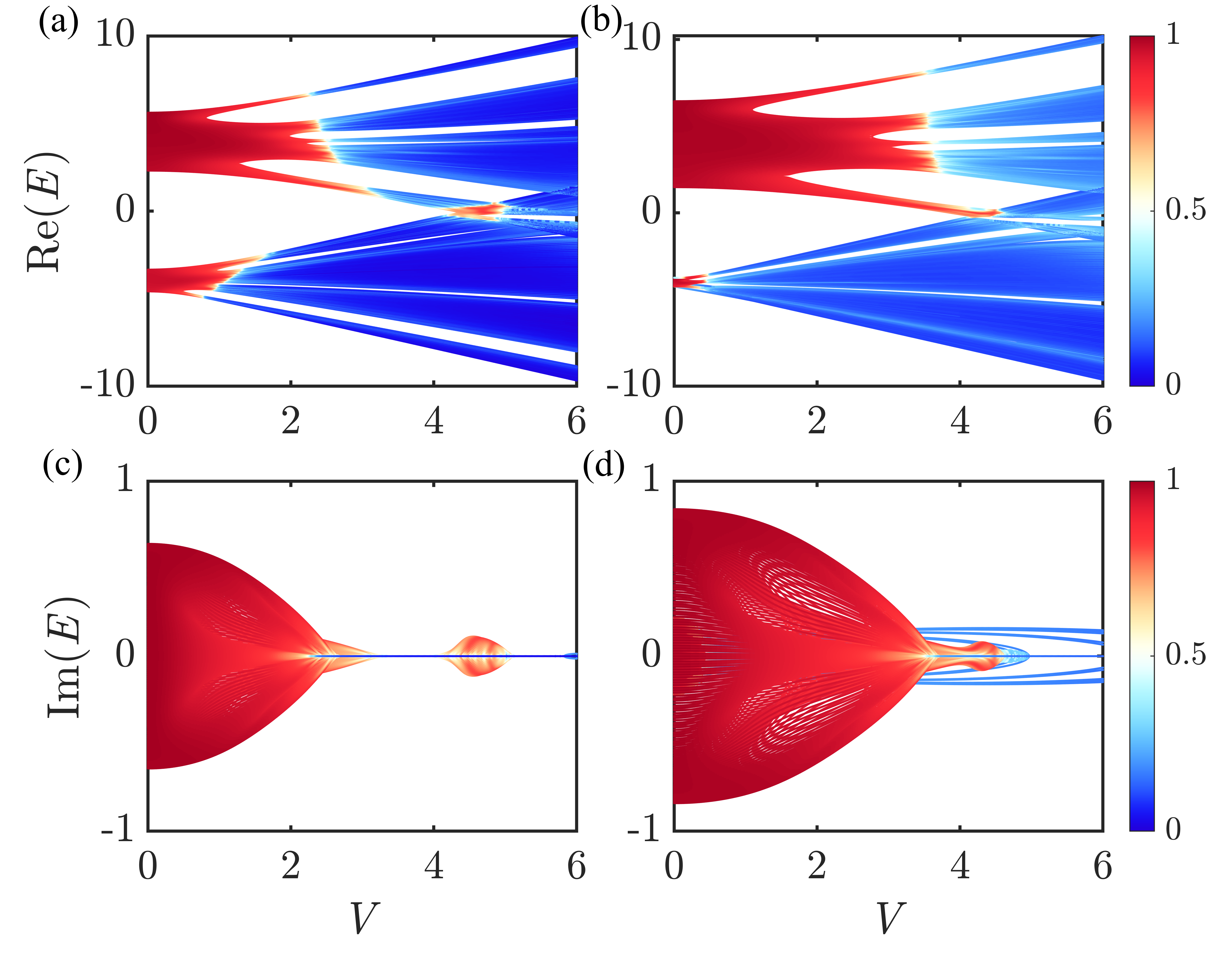}
	\caption{The real and imaginary parts of energy spectrum versus the strength of quasiperiodic potential $V$ for the system with $t_1=1,t_2=4, \alpha=0.25$. Here
             (a,c) $t_3=0.25$ and (b,d) $t_3=0.65$, respectively. The colormap shows $\Gamma^i$ associated with $i$th eigenstate.}
	\label{Fig5}
\end{figure}
The reentrant localization transition could also exist in the non-Hermitian system by adjusting the strength of asymmetric hopping between different lattices. In this section, we investigate a non-Hermitian Hamiltonian M2 corresponding to a one-dimensional generalized AAH quasiperiodic lattice with asymmetric coupling as shown in Table \ref{I}.

To identify the effect of long-range hopping on the reentrant localization transition, we first calculate the real and imaginary parts of the eigenenergies and their associated $\Gamma^i$ as functions of $V$. As shown in Fig. \ref{Fig5}(a), there exist two separate intermediate regions with mobility edges where the extended states and localized states coexist. Similar to the Hermitian case, some already localized states become extended again and then localized finally, a phenomenon known as reentrant localization. Combined with the imaginary part of the energy spectrum in Fig. \ref{Fig5}(c), it is found that the complex to real, real to complex, and complex to real transitions in the spectrum all coincide with the localization transitions \cite{wu2021non}. In addition, the imaginary parts of the eigenenergies vanish completely within each localized phase. Moreover, the introduction of $t_3$ will also affect the chiral symmetry and modified the eigenenergy spectrum as shown in Fig. \ref{Fig5}(b). By analyzing the energy spectrum with the encoded $\eta$, more extended states tend to be distributed in the upper band and two non-adjacent red extended regions in the upper band merge together. As a consequence, the reentrant feature vanishes, and only one intermediate phase with the mobility edge is preserved in Fig. \ref{Fig5}(b). From the imaginary part of the energy spectrum in Fig. \ref{Fig5}(d), we can know that there are no real to complex and complex to real transitions anymore.

\begin{figure}[]
	\centering
	\includegraphics[trim=5mm 0mm 0mm 0mm,scale=0.38]{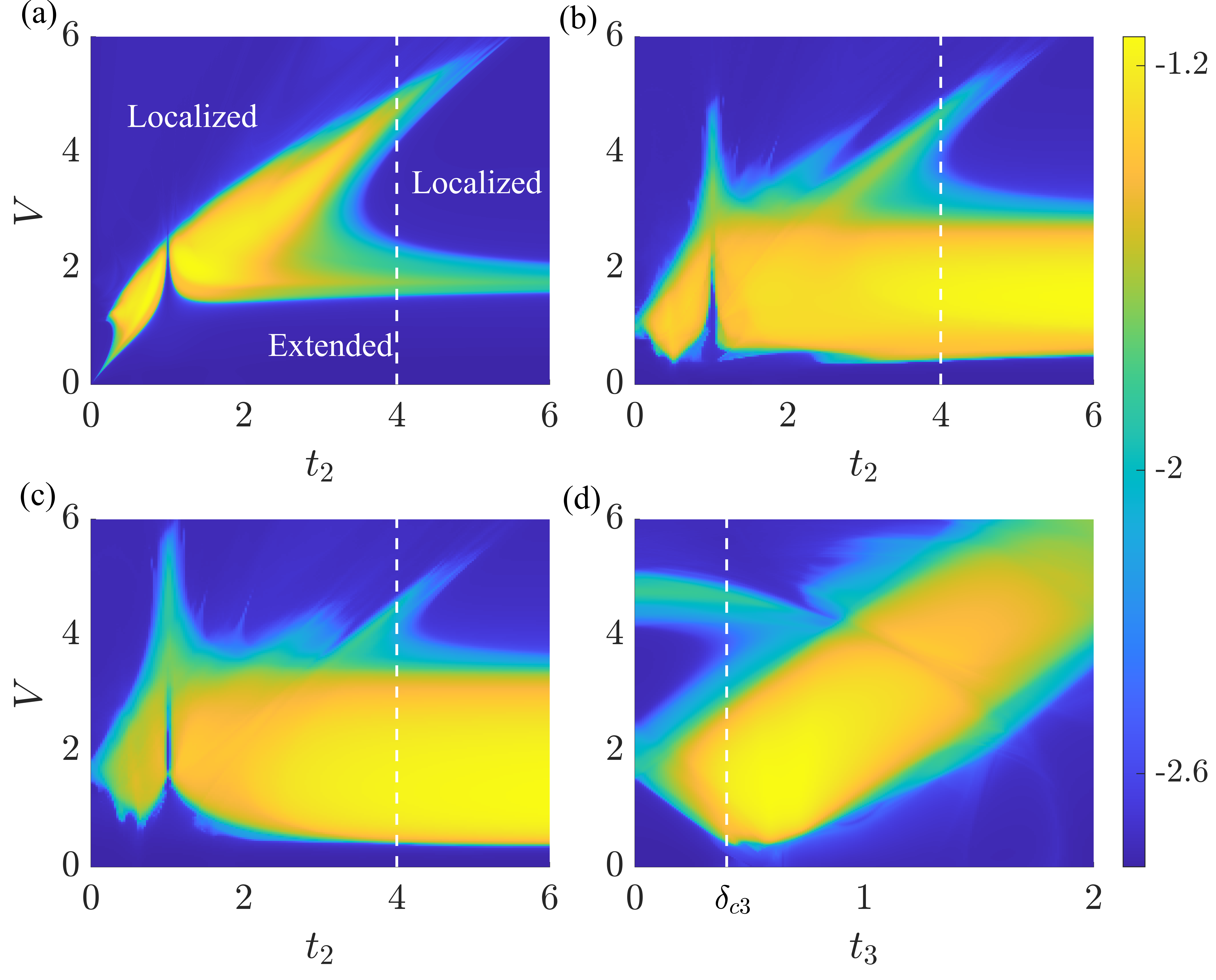}
	\caption{Phase diagram of the system in $V$ and $t_2$ plane with non-Hermitian staggered quasiperiodic disorder for (a) $t_3=0$, (b) $t_3=0.4$, (c) $t_3=0.6$. (d) Phase diagrams of the non-Hermitian system in $V$ and $t_3$ plane for $t_2=4$. Colorbar indicates different value of $\eta$, the extended and localized phases are represented by dark blue regions, and the brilliant yellow region shows the intermediate phase with mobility edges. Here $\alpha=0.25, t_1=1$.}
	\label{Fig6}
\end{figure}

To investigate the localization behavior of the system, we further investigate the quantity $\eta$ as a phase diagram in Fig. \ref{Fig6}. With the presence of asymmetric hopping, the region of SPME is expanded. Similar to the Hermitian case, there are also two critical regions when $t_{2}$ is between $3.56$ and $5.4$, which exactly indicates the existence of the reentrant localization. More importantly, when the system is subjected to appropriate NNNH strength (as shown in Fig. \ref{Fig6}(b, c)), this reentrant feature is absent with $t_2=4$ while $t_3 \ge 0.4$. To verify the role of NNNH in non-Hermitian systems, we study the phase diagram of $\eta$ versus $V$ and $t_{3}$ by fixing $t_2=4$. In Fig. \ref{Fig6}(d), two intermediate regions merged together when $t_3$ exceed a critical point $\delta_{c3} \approx$ 0.4, this is consistent with the energy spectrum calculations presented in Fig. \ref{Fig5}. In analogy to the Hermitian system, the introduction of long-range hopping would drive the reentrant localization to disappear when asymmetric hopping is present.

\begin{figure}[]
	\centering
	\includegraphics[trim=5mm 0mm 0mm 0mm,scale=0.43]{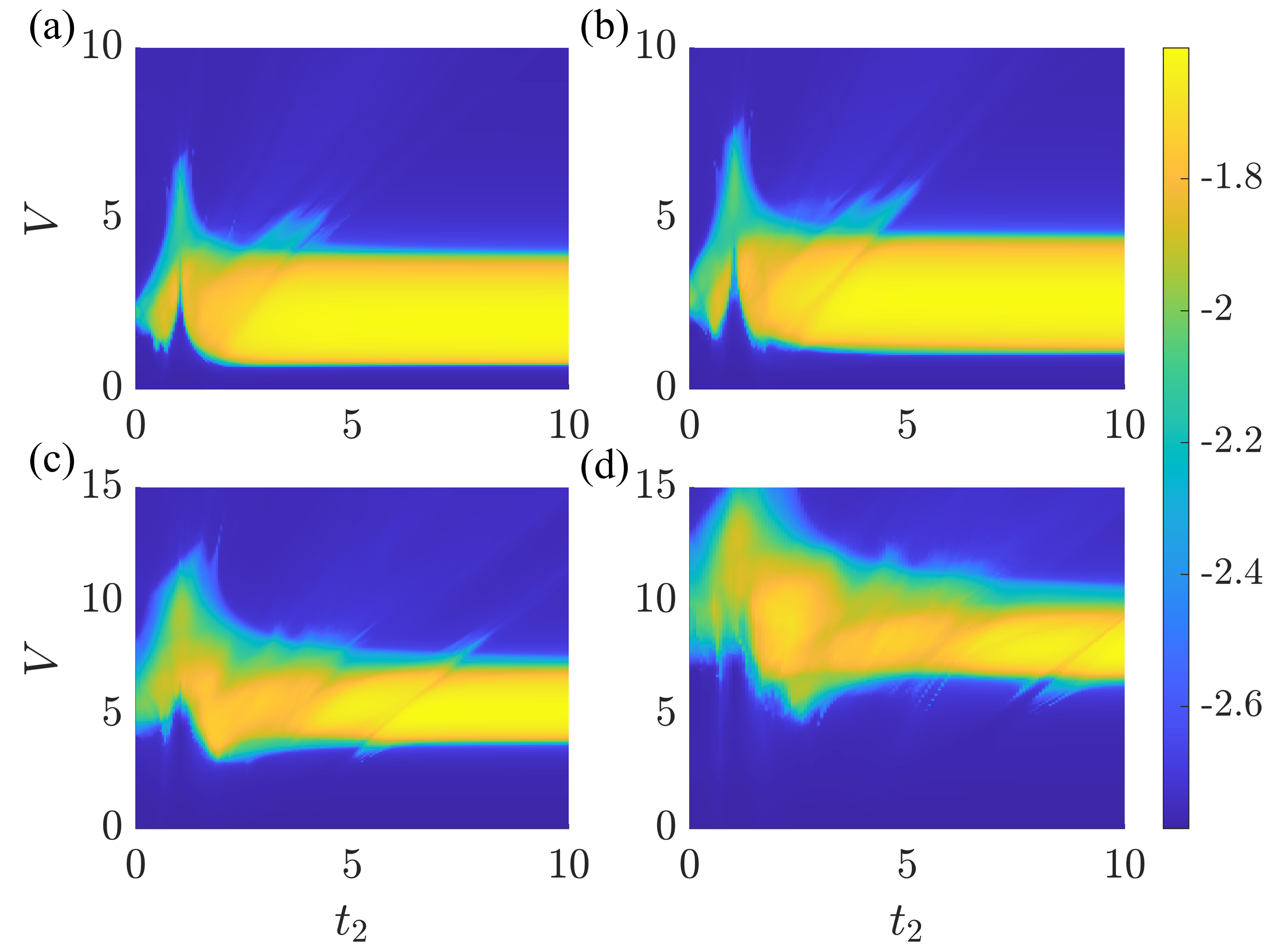}
	\caption{Phase diagrams in $V-t_2$ plane for different strength of NNNH. Here (a) $t_ 3=0.8$, (b) $t_ 3=1$, (c) $t_ 3=2$, (d) $t_3=3$, respectively. The colorbar indicate values of $\eta$, other parameters are the same as those in Fig. \ref{Fig6}(a).}
	\label{Fig7}
\end{figure}
As illustrated in Fig. \ref{Fig6}(a), as compared to the phase diagram presented in Fig. \ref{Fig3}(a) in Hermitian system, the parameter interval of $t_2$ for which two critical regions coexist is enlarged due to the introduction of asymmetric hopping. However, phase diagram in Fig. \ref{Fig6}(b, c) only shows the disappearance of the second critical region in a fixed $t_3$, but if the strength of $t_3$ continues to increase, this phenomenon will reappear. For this reason, to find out whether this phenomenon is completely disappeared or not in the $V-t_2$ plane when the NNNH is present, we plot the phase diagram for different NNNH strengths in Fig. \ref{Fig7}. From Fig. \ref{Fig7}(a, b), we find for $t_3=0.8$ and $t_3=1$, even though the second critical region is gradually shrinking, but it still exists in the presence of NNNH. However, if the strength of NNNH is large enough such as $t_3=2$ and $t_3=3$, the reentrant localization eventually disappears as shown in Fig. \ref{Fig7}(c, d).

\subsection{M3: non-Hermitian Hamiltonian with complex phase factor} \label{Sec:III C}
In general, the non-Hermiticity can also be achieved by introducing the complex on-site potential or asymmetric coupling in a non-Hermitian system. In this section, the non-Hermiticity of the Hamiltonian is generated by a complex on-site potential and satisfies $V_{n}=V_{-n}^{*}$ \cite{jiang2021mobility} and possesses the $\mathcal{PT}$ symmetry, which shown in M3 of the Table \ref{I}.

\begin{figure}[]
	\centering
	\includegraphics[trim=3mm 0mm 0mm 0mm,scale=0.33]{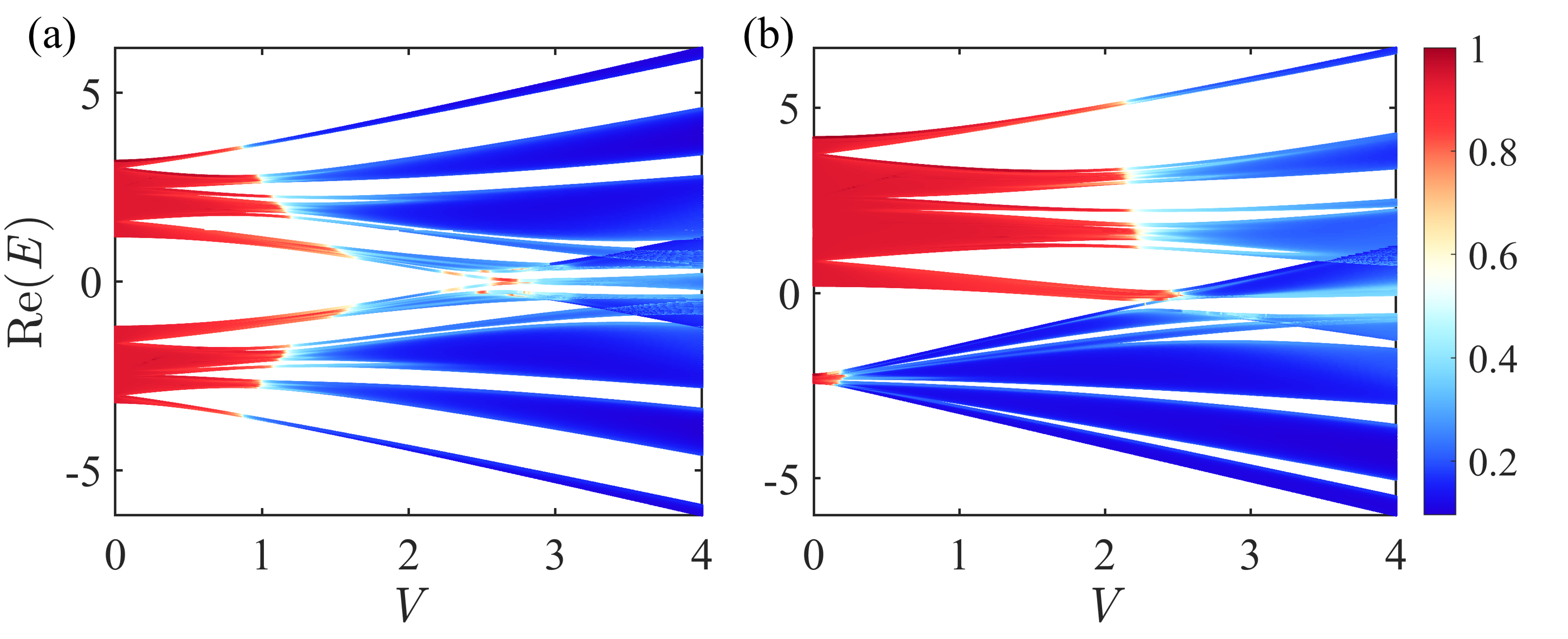}
	\caption{The real parts of the eigenvalue spectra versus $V$ for the system with (a)$t_{3} = 0$, (b)$t_{3} = 0.5$. Here other model parameters are $\Delta=0.05$ and $t=2.2$. The colorbar indicate values of $\Gamma^{i}$.}
	\label{Fig8}
\end{figure}
The localization feature can be encoded in the energy spectrum with the corresponding $\Gamma^{i}$. In Fig. \ref{Fig8}, we present the real parts of the eigenvalue spectrum while the quasi-periodic potential has a complex phase factor. Similar to the two cases mentioned above, in Fig. \ref{Fig8} (a), the system will undergo a series of reentrant localization transitions without NNNH. However, by introducing the NNNH shown in Fig. \ref{Fig8} (b), the band structure and corresponding localized(extended) states distribution will be modified, and the dimerization of the system will be weakened, resulting in the disappearance of the reentrant feature with the participation of NNNH.

\begin{figure}[]
	\centering
	\includegraphics[trim=3mm 0mm 0mm 0mm,scale=0.38]{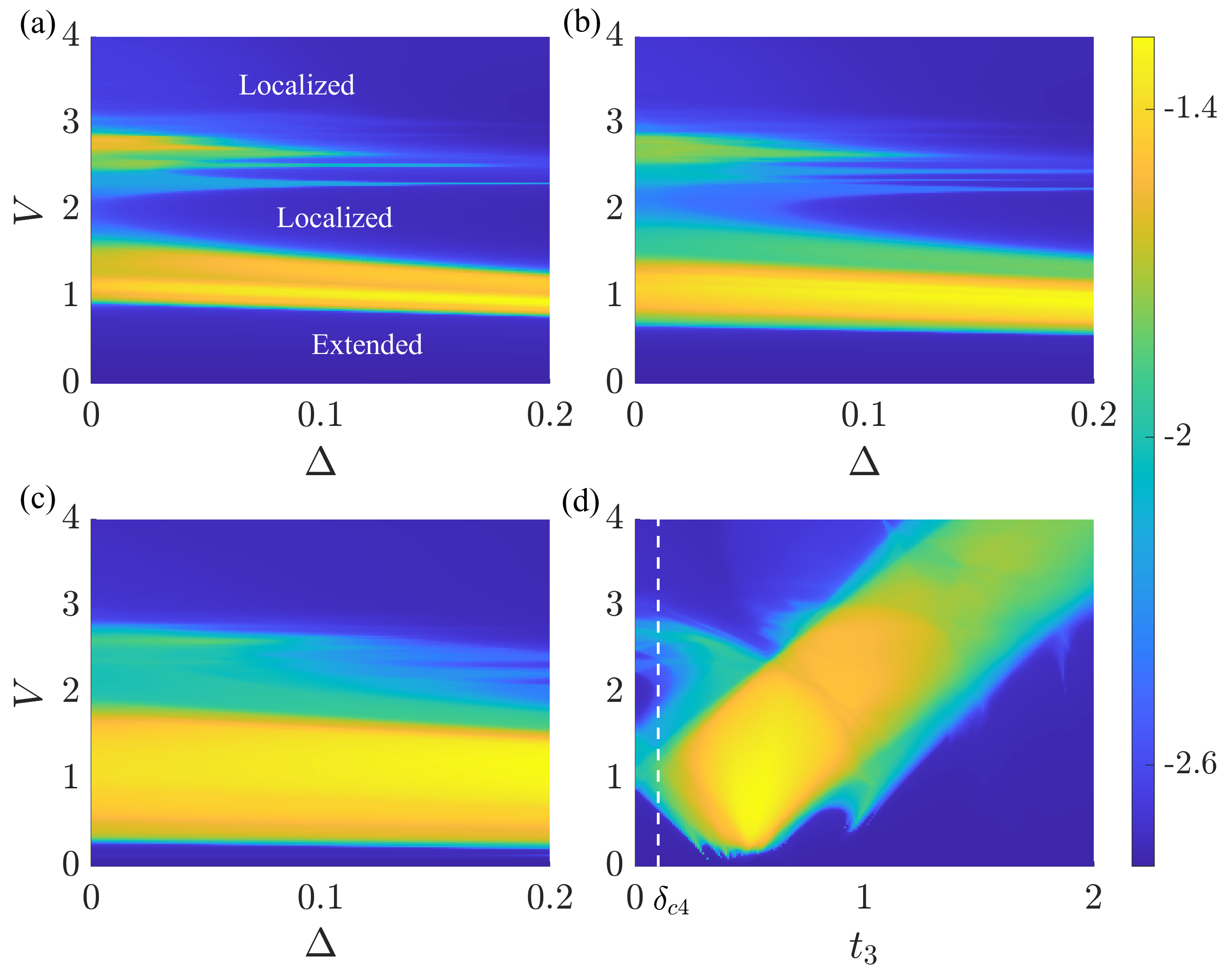}
	\caption{Phase diagram of the non-Hermitian system in $V$ and $\Delta$ plane for (a) $t_3=0$, (b) $t_3=0.1$, (c) $t_3=0.3$, respectively. (d) Phase diagram in $V$ and $t_3$ plane for $\Delta=0.05$. The colorbar indicates different values of $\eta$. Here $t_{1}=1,t_{2}=2.2$.}
	\label{Fig9}
\end{figure}
In order to reveal the localization behavior of the system more obviously. As illustrated in Fig. \ref{Fig9}(a), we present the phase diagram in the $V-\Delta$ plane using the numerical calculation of $\eta$. At first, when $\Delta \le 0.036$ and without NNNH, there are two critical regions that represent the reentrant localization. But if complex phase factor $\Delta$ continues to increase, there exist the multiple intermediate phases for $\Delta=[0.036, 0.18]$ and the system undergoes multiple localization transitions. However, while the strength of $\Delta$ is large enough, the top branch eventually decays. Finally, only one intermediate phase will remain, and the reentrant property will be lost.
In summary, the introduction of long-range hopping can eliminate the reentrant phenomenon in above cases. To investigate whether the introduction of NNNH would also kill this feature with the complex phase factor. In Fig. \ref{Fig9}(b,c), we present the phase diagram in the $V-\Delta$ plane for different NNNH strength. We observe that two critical regions merged and the reentrant localization transition vanishes as the increase of NNNH strength. Similar to the above result in Fig. \ref{Fig3}(d) and Fig. \ref{Fig6}(d), after a critical point $\delta_{c4} \simeq  0.1$, the reentrant feature at an exact complex phase factor will also vanish by the increase of $t_3$, as shown in Fig. \ref{Fig9}(d). As a matter of fact, due to the introduction of the NNNH, the reentrant property will also fade away while the Hamiltonian is hosting a complex on-site potential. From what we have mentioned above, we can conclude that no matter what kind of case the system is in the Table \ref{I}, the introduction of long-range hopping will finally eliminate reentrant localization in the different phase diagrams.

\begin{figure}[]
	\centering
	\includegraphics[trim=3mm 0mm 0mm 0mm,scale=0.31]{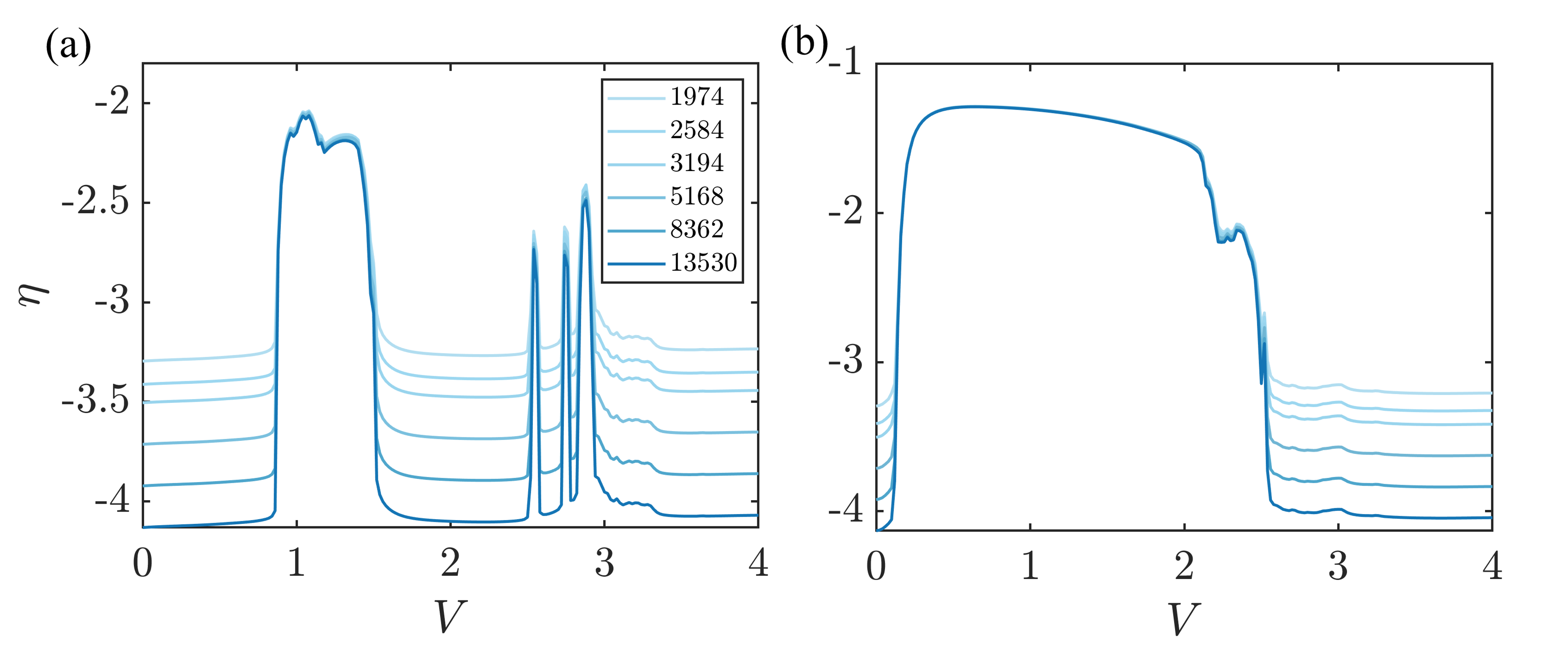}
	\caption{The $\eta$ plot as a function of $V$ for $L= 1974,2584,3194,5158,8362,13530$ from top to bottom for (a) $t_3=0$, (b)$t_3=0.5$, respectively. Here other model parameters are $\Delta=0.05$ and $t=2.2$.}
	\label{Fig10}
\end{figure}
To confirm the multiple reentrant localization and rule out the finite size effects in Fig. \ref{Fig8} and Fig. \ref{Fig9}(a). We compute the $\eta$ for different system sizes with fixed $\Delta$. As the system grows larger, four intermediate regions become particularly distinct, as shown in Fig \ref{Fig10}(a), which definitely establishes the stability of this behavior in the absence of NNNH. To eliminate the multiple reentrant behaviors, we set $t_3=0.5$ in Fig \ref{Fig10}(b), which is identical to the prior approach. As expected, with the presence of NNNH, four intermediate phases merged together, and finally, only one intermediate phase survives. Moreover, computations under different system sizes support the stability of this elimination behavior.

\section{Discussion and conclusion} \label{Sec:IV}
The reentrant physics of the non-Hermitian AAH model in Eq. (\ref{eqI}) can be simulated by electric circuits\cite{PhysRevResearch.2.033052}, which are proven to be a powerful platform for investigating non-Hermitian and/or topological phases\cite{Ashida_2020}. Moreover, negative impedance converters with current inversion (INIC)\cite{PhysRevLett.122.247702, helbig2020generalized} can realize the asymmetrical hopping amplitudes, and complex on-site potentials can be simulated by grounding nodes with proper resistors\cite{PhysRevA.84.040101, Rafi-Ul-Islam_2021}. By measuring two-node impedances, the energy spectrum could be obtained\cite{PhysRevB.101.020201}.

In conclusion, we have studied the localization transition in a dimerized lattice with the staggered quasiperiodic disorder for three different Hamiltonian cases, which is illustrated in Table \ref{I}. We discovered that no matter what kind of case the system is, the introduction of long-range hopping will finally eliminate reentrant localization in different phase diagrams, beyond the critical values of $t_3$, two separate intermediate regions tend to merge together, and the result is shown as M1-M3 cases in the main text. This can be attributed to the fact that the increase of NNNH weakens the dimerization of the SSH chain, thus destroying the competition between dimerization and quasi-periodic disorder, which results in the disappearance of reentrant localization transition and confirms the irreplaceable importance of dimerization to generate the reentrant phenomenon. On the other hand, while the non-Hermiticity is introduced by the complex on-site potential illustrated in M3, the increasing non-Hermitian parameter $\Delta$ will also significantly remove the reentrant localization transition in a fixed hopping dimerization. All in all, the reentrant phenomenon would be eliminated not only due to the introduction of long-range hopping but also owing to the increasing complex phase factor. We confirm this finding by examining the participation ratios, eigenspectra, and different phase diagrams in the system.

$${\bf ACKNOWLEDGMENTS}$$

We gratefully acknowledge the support from the National Natural Science Foundation of China (Grant No. 12074230, 12174231, 11974355, 12147215), the Fund for Shanxi ``1331 Project", Shanxi Province 100-Plan Talent Program, Fundamental Research Program of Shanxi Province through 202103021222001. This research was partially conducted using the High Performance Computer of Shanxi University.

\bigskip

\noindent{$^{*}$chenjun@sxu.edu.cn}

\noindent{$^{\dagger}$zhanglei@sxu.edu.cn}

\bibliography{ref}

\end{document}